\documentclass[12pt]{article}

\usepackage{amsmath}
\usepackage{amssymb}
\usepackage[utf8]{inputenc}
\usepackage[super, comma,sort&compress]{natbib}
\usepackage{float}
\usepackage[T1]{fontenc}
\usepackage{graphicx}
\usepackage{mathtools}
\usepackage{txfonts}
\usepackage[normalem]{ulem}
\usepackage{wasysym}
\usepackage[svgnames]{xcolor}
\usepackage[paperheight=29.7cm,paperwidth=21cm,left=2.54cm,right=2.54cm,top=2.54cm,bottom=2.54cm]{geometry}
\usepackage[hidelinks]{hyperref}
\usepackage{indentfirst}
\usepackage{xcolor}
\begin{document}

\textbf{\begin{center}Observation of three-state nematicity and domain evolution in atomically-thin antiferromagnetic NiPS\textsubscript{3}\end{center}}

Qishuo Tan\textsuperscript{1\dag}, Connor A. Occhialini\textsuperscript{2\dag}, Hongze Gao\textsuperscript{1}, Jiaruo Li\textsuperscript{2}, Hikari Kitadai\textsuperscript{1}, Riccardo Comin\textsuperscript{2*}, and Xi Ling\textsuperscript{1, 3, 4*}
\vspace{1\baselineskip}

\textsuperscript{1} Department of Chemistry, Boston University, Boston, Massachusetts, 02215, USA. 

\textsuperscript{2} Department of Physics, Massachusetts Institute of Technology, Cambridge, Massachusetts, 02139, USA. 

\textsuperscript{3} Division of Materials Science and Engineering, Boston University, Boston, Massachusetts, 02215, USA. 

\textsuperscript{4} The Photonics Center, Boston University, Boston, Massachusetts, 02215, USA.
\vspace{1\baselineskip}

\text{\dag These authors contributed equally to this work.}

\textsuperscript{}*To whom the correspondence should be addressed. Email address: \href{mailto:rcomin@mit.edu}{\uline{\textcolor[HTML]{0563C1}{rcomin@mit.edu}}};\\ \href{mailto:xiling@bu.edu}{\uline{\textcolor[HTML]{0563C1}{xiling@bu.edu}}}

\vspace{1\baselineskip}
\vspace{1\baselineskip}

\newpage

\section*{Abstract}
\textbf{Nickel phosphorus trisulfide (NiPS\textsubscript{3}), a van der Waals (vdW) 2D antiferromagnet, has captivated enormous attention for its intriguing physics in recent years. However, despite its fundamental importance in physics of magnetism and promising potential for technological applications, the study of magnetic domains in NiPS\textsubscript{3} down to atomically thin is still lacking. Here, we report the layer-dependent magnetic characteristics and magnetic domains within antiferromagnetic NiPS\textsubscript{3} by employing linear dichroism (LD) combined with polarized microscopy, spin-correlated photoluminescence (PL), and Raman spectroscopy. Our results reveal the existence of the paramagnetic-to-antiferromagnetic phase transition in bulk to bilayer NiPS\textsubscript{3} with stronger spin fluctuation in thinner NiPS\textsubscript{3}. Furthermore, our study identifies three distinct antiferromagnetic domains within atomically-thin NiPS\textsubscript{3} and captures the thermally-activated domain evolution. Our findings provide crucial insights for the development of antiferromagnetic spintronics and related technologies.}

\section*{Main}
Antiferromagnets (AFMs) are materials in which magnetic moments align in long-range order yet with zero net magnetic moment below a critical transition temperature. Historically overshadowed by the prominence of ferromagnets, AFMs were once labelled as "interesting and useless" for quite a long time\cite{Neel_Nobel}. The challenges were how to detect the antiferromagnetic order, and its control required extremely large magnetic field. Nowadays, the advancement of technology has ushered in new methodologies for studying AFMs, including those leveraging light-matter interactions\cite{MagneticReview, MagneticReview2}. Revealed through the combination of conventional and new approaches, the intrinsic strengths of AFMs such as terahertz exchanging, multilevel states and absence of stray fields, have positioned them as promising candidates for the next-generation spintronics\cite{AFM_spintronics, AFM_optospintronics}. As a crucial factor affecting the performance of applications, antiferromagnetic domain has thus garnered considerable attention to be investigated on\cite{seeing_AFM_domain, AFM_domainwall_mechanics, AFM_sharpdomain}, especially in two dimensional (2D) region\cite{MnPS3domain, FePS3domain1, MnPSe3domain}.

\hspace{1em}Nickel phosphorus trisulfide (NiPS\textsubscript{3}), as an emerging van der Waals (vdW) 2D antiferromagnetic material from the family of transition metal trichalcogenides, offers an exceptional platform for investigating the electronic and magnetic properties of strongly correlated 2D systems, such as phonon-magnon coupling\cite{phonon-magnon, NiPS3magnon}, spin-correlated emission\cite{spin-correlated1, spin-correlated2,spin-correlated3}, and \textit{d} electron scattering\cite{electronicRaman} or transition\cite{dd_dark, ddemission}. However, despite these interesting spin-correlated phenomena indicate enormous promise for AFMs-based spintronics, the investigation of magnetic domain and its evolution in NiPS\textsubscript{3} especially at atomically-thin regime is still in demand.

\hspace{1em}In this study, we use magneto-optical linear dichroism (LD) response to probe the thickness-dependent antiferromagnetism and the magnetic domains in NiPS\textsubscript{3} down to atomically-thin layers combined with polarized microscopy, PL, and Raman spectroscopy. Our temperature-dependent LD measurements across samples spanning from two layers (2 L) to five layers (5 L) reveals a trend of decreasing magnetic phase transition temperature and stronger spin fluctuation in thinner NiPS\textsubscript{3}. Notably, no LD signals are detected from monolayer NiPS\textsubscript{3}, indicating the absence of long-range magnetic ordering. Moreover, three-state nematicity is observed in atomically thin NiPS\textsubscript{3} by the angular-dependent LD measurements, further supported by characterizations through polarized exciton emission and Raman spectroscopy. The evolution of magnetic domains with temperature increasing in a 2 L NiPS\textsubscript{3} sample is also captured, illustrating a process of thermally activated domain wall motion.

\subsection*{Linear dichroism and magnetic domain}
Bulk NiPS\textsubscript{3} has a monoclinic stacking configuration within the space group \textit{C}2/\textit{m} with weak interlayer coupling\cite{phonon-magnon, NiPS3weakinterlayer}. As shown in Fig. 1a, the Ni atoms arranged in a hexagonal lattice establish an in-plane zigzag antiferromagnetic order within each layer at temperature below its N\'eel temperature (${T_{\mathrm{N}}}$). Previous neutron scattering measurements on bulk NiPS\textsubscript{3} show that the spins of Ni atoms align either parallel or anti-parallel along the \textit{a} axis in the \textit{ab} plane, featuring a slight out-of-plane component of approximately \textasciitilde7$^{\circ}$\cite{NiPS3AFM}. Interlayer spins are coupled ferromagnetically. As for NiPS\textsubscript{3} flakes, the results of polarization-dependent second harmonic generation (SHG) show sixfold pattern at both above and below ${T_{\mathrm{N}}}$\cite{MnPS3_SHG}, suggesting that they persist approximate C\textsubscript{3} rotational symmetry as monolayer NiPS\textsubscript{3} has. Thus, as the in-plane zigzag antiferromagnetic order emerges and breaks the approximate C\textsubscript{3} symmetry, there would be three possible directions related by 120$^{\circ}$ to each other as depicted in Fig. 1b, which indicates the formation of three-state Potts nematic order ($\mathbb{Z}_3$)\cite{Potts}. The similar persistence of approximate C\textsubscript{3} symmetry and symmetry breaking due to zigzag antiferromagnetic order have also been observed in monoclinic FePS\textsubscript{3}\cite{FePS3domain2}. To investigate the intriguing magnetic properties in NiPS\textsubscript{3} flakes, we mechanically exfoliated them from the bulk single crystal on silicon substrates (e.g., Fig. S1a). The spin-correlated sharp PL peak of 1.476 eV, and the characteristic Raman peaks are observed (Fig. S1b-c), indicating that the samples are in high-quality. 

\hspace{1em}The measurement of magnetic linear dichroism (LD) whether employing laser or X-ray light as an excitation, offers a potent technique for probing the symmetry breaking in 2D materials\cite{FePS3domain1, FePS3cavity, NiI2, FeSeXLD, FePS3XLD}. For zig-zag AFMs, the anisotropic absorption of linearly polarized light parallel and perpendicular to the N\'eel vector allows for the determination of the magnetic orientation. Figure 1c illustrates the experimental setup for the reflective LD measurements, with more details in the Methods section. An excitation laser light with tunable wavelength, modulated by a photo-elastic modulator (PEM), is directed onto NiPS\textsubscript{3} samples and the reflection signals are collected by a photodiode and subsequently analyzed through a lock-in amplifier. Figure 1d shows the temperature-dependent LD spectra of the thick NiPS\textsubscript{3} sample (Fig. S1a), with prominent signals arising while the energies are in resonance with the charge-transfer exciton\cite{NiPS3charge-spin_correlation} at around 1.9-2.1 eV and the spin-correlated exciton\cite{spin-correlated3} at around 1.5 eV. With temperature increasing, these signals diminish until totally vanish above ${T_{\mathrm{N}}}$. We extract the LD signals at the energy of 2.05 eV and plot the temperature dependence in Fig. 1e. The temperature-dependent LD intensity is fitted using the formula

\centerline{$\mathrm{LD}(T) \sim\left|1-\frac{T}{T_{\mathrm{N}}}\right|^{2\beta_{\mathrm{LD}}}$ (1),}

where ${T_{\mathrm{N}}}$ is the transition temperature, and $\beta_{\mathrm{LD}}$ is the critical exponent describing the evolution of the order parameter through the magnetic phase transition\cite{criticalexponent,criticalexponent2}. Thick NiPS\textsubscript{3} shows a ${T_{\mathrm{N}}}$ of 157 K and a $\beta_{\mathrm{LD}}$ of 0.25.

\hspace{1em}We further use a cross-polarized microscopy to characterize the magnetic domains, where an incident linearly polarized light impinges on the sample and the reflected light is detected. Due to the birefringence effect induced by the magnetization vector, the polarization of the reflected light would depend on magnetic domains, resulting in a contrast difference for magnetic domains by detuning the incident and reflected light polarization\cite{PolarizedMicroscopy}. Figure 1f shows the polarized microscopy image of the thick NiPS\textsubscript{3} sample measured at 5 K, wherein the contrast difference signifies the presence of two distinct domains, i.e. the major domain\_1 outlined by green dashed lines, and the smaller domain\_2 outlined by blue dashed lines. The alteration in contrast sign of these two domains when changing the detuning angle indicates their magnetic origin (Fig. S2a). In addition, as we increase the temperature, the domains vanish when \textit{T} is higher than ${T_{\mathrm{N}}}$ of 157 K (Fig. S2b). The coexistence of antiferromagnetic domains is further substantiated through LD mapping within the same area (Fig. 1g), where the same two magnetic domains are observed as those from the polarized microscopy measurements. To probe the direction of the N\'eel vector, we measure the angular-dependent LD, with the incident laser polarization rotated by a half-wave plate, on each mono-domain region. Since the light absorption parallel to the N\'eel vector is stronger than that perpendicular to the N\'eel vector for NiPS\textsubscript{3}\cite{spin-correlated3}, the direction of the  negative maxima of the reflective LD results showcases the direction of the N\'eel vector, i.e., the orientation of the zig-zag antiferromagnetic ordering. From the angular-dependent LD results shown in Fig. 1h, the N\'eel vector of domain\_1 is along 0$^{\circ}$, but it is along 60$^{\circ}$ (-120$^{\circ}$) for domain\_2. This angular discrepancy of 120$^{\circ}$ for the antiferromagnetic ordering matches with the three-state nematicity shown in Fig. 1b. Furthermore, given the report that the spin-correlated PL is perpendicular to the N\'eel vector\cite{spin-correlated3}, we also conduct PL mapping and angular-dependent PL measurements on the same area (Fig. S3), yielding outcomes that align with the aforementioned techniques. These two domains related by 120$^{\circ}$ in their N\'eel vectors implies the existence of the third domain which are observed in atomically-thin samples presented in the following. 

\subsection*{Layer-dependent antiferromagnetism}
To investigate the magnetism and domains further in atomically thin samples, we extend the LD measurements to NiPS\textsubscript{3} flakes from 1 L to 5 L. Figure 2a presents the optical image of the NiPS\textsubscript{3} flakes, showing areas ranging from 1 L to 5 L that are used for the measurements. The optical contrast extracted from the red channel of distinct regions is shown in Fig. 2b, facilitating the determination of the layer number\cite{NiPS3optical}. Figure 2c shows the layer-dependent LD results measured at 1.75 K, revealing robust signals from 2 L to 5 L, whereas no signals from the monolayer are observed. The layer-dependent LD intensity is summarized in Fig. 2e, showing a nonlinear increase with the increase of the layer number. We further performed the temperature-dependent LD measurements on these regions (Fig. 2d). With the increase of the temperature, a remarkable decrease of the LD intensity is observed for 2 L to 5 L samples, which eventually vanishes above the N\'eel temperature. For monolayer sample, no visible signal and phase transition are observed. The result demonstrates unambiguous phase transition for all thin flakes except the monolayer. The fitted layer-dependent values of ${T_{\mathrm{N}}}$ and $\beta_{\mathrm{LD}}$ by equation (1) are summarized in Fig. 2f and 2g, respectively. Evidently, the transition temperature decreases from 154 K for the 5 L flake to 121 K for the 2 L flake. This conspicuous dimensional effect of lower ${T_{\mathrm{N}}}$ for thinner sample is also observed in other 2D antiferromagnetic materials such as MnPS\textsubscript{3}\cite{MnPS3layer1, MnPS3layer2, MnPS3layer3}, FePS\textsubscript{3}\cite{FePS3layer1, FePS3layer2,FePS3layer4}, MnPSe\textsubscript{3}\cite{MnPSe3domain, MnPSe3layer1}. In addition, the critical exponent $\beta_{\mathrm{LD}}$ increases from 0.25 for the 5 L flake to 0.35 for the 2 L flake. The $\beta_{\mathrm{LD}}$ of 0.25 for 5 L and thick NiPS\textsubscript{3} samples aligns well with previous report\cite{spin-correlated3} and reflects the universal signature of 2D XY magnetic system\cite{2DXY}. However, as NiPS\textsubscript{3} becomes thinner, larger critical exponent is observed which indicates sharper decrease of the magnetization. This may be induced by the weaker easy-plane-like anisotropy, and implies stronger spin fluctuation in all three dimension for thinner NiPS\textsubscript{3}\cite{phonon-magnon}.

\hspace{1em} More LD measurements in the range of 1.4 to 2.6 eV are preformed on 1 L to 4 L samples (Fig. S4). Clear LD signals within the energy range of 1.7 eV and 2.3 eV are observed in the 2 L and thicker samples, which correspond to the energy of charge-transfer exciton. Furthermore, the extracted LD peaks show a blue shift in thinner samples, resulting from the quantum confinement effect. For monolayer NiPS\textsubscript{3}, LD signals are not observed in the energy range we measured for temperature in the range of 2 K to 150 K due to the suppression of the magnetic ordering\cite{phonon-magnon}. We also perform the Raman spectroscopy to support the suppression of magnetic ordering in monolayer NiPS\textsubscript{3} and discuss it in Supplementary Materials (Fig. S5).

\subsection*{Observation of three-state nematicity}
We then apply the angular LD measurements to study the magnetic domains of atomically-thin NiPS\textsubscript{3} samples. Figure 3a shows the optical image of a four-layer NiPS\textsubscript{3} flake with the edges delineated by black dashed lines. LD mappings with the incident laser polarization at 0$^{\circ}$, 60$^{\circ}$, 90$^{\circ}$, 120$^{\circ}$, 150$^{\circ}$ and 210$^{\circ}$ are performed at 4 K (Fig. 3b-d and Fig. S6), where the LD signal flip from negative maximum to positive maximum when rotating the incident laser polarization by 90$^{\circ}$, indicating the existence of the magnetic domain. For better visualization, each of the mono-domain regions is outlined in green, blue, and red in the corresponding Fig. 3b, c, and d. The angular-dependent LD of each highlighted mono-domains is measured and shown in Fig. 3e. It clearly shows that the N\'eel vector aligns at 0$^{\circ}$, 60$^{\circ}$ (-120$^{\circ}$), and 120$^{\circ}$ for domain\_1, domain\_2, and domain\_3, respectively, demonstrating the observation of the three-state nematicity as represented by the cartoons in Fig. 3g. Interestingly, the edge of the sample shows a new LD pattern different from any domain of the inner region (Fig. S6b) that may result from local structure distortion\cite{MoS2_edge, BFanomalous_phonon}, which falls beyond the scope of this study. We further perform the angular-dependent PL measurements on the three observed domains. Figure 3f shows the intensity polar plots of the sharp PL emission at 1.476 eV for each magnetic domain, presenting signals reach a maximum when the collection polarization is along 90$^{\circ}$, 150$^{\circ}$, and 210$^{\circ}$ for the three domains, respectively. As the sharp PL emission is spin-correlated and the polarization of the PL emission is perpendicular to the N\'eel vector\cite{spin-correlated3, NiPS3domain}, the result suggests that the N\'eel vector of domain\_1, domain\_2 and domain\_3 is along 0$^{\circ}$, 60$^{\circ}$ (-120$^{\circ}$), and 120$^{\circ}$, respectively, which is consistent with the results obtained from the LD measurements, further confirming the existence of the three-state nematicity. As Raman spectroscopy was recently reported to be able to probe the magnetic order and the crystal axis\cite{NiPS3domain}, we further conduct the polarized Raman measurements on the same flake as well as the adjacent thicker region (Fig. S7). The result reveals that the major domain (i.e. domain\_1) of the 4 L NiPS\textsubscript{3} possesses the antiferromagnetic order along the \textit{a}-axis. Similarly, the three-state nematicity is also observed in a thinner 2-3 L flake by conducting LD mapping and Raman measurements as shown in Fig. S8-10. 

\subsection*{Thermally-activated domain evolution}
To delve into the magnetic domain evolution, we perform the LD mapping on the 2-3 L NiPS\textsubscript{3} sample with temperature changing in the range of 2 K to 175 K. Figure 4a showcases the representative domain evolution, with additional results presented in Fig. S11. All three domains are observed in both the 2 L and 3 L regions at 2 K. For the 2 L flake, the major domain is domain\_2, accompanied by a small domain\_1 situated in the middle, while domain\_3 remains pinned near the right edge. With increasing temperature, the small domain\_1 within the 2 L region starts to grow from its edge, extending towards the right section of the flake. As the temperature reaches 105 K, the entire right section of the 2 L region transitions into mono-domain\_1 completely. When the temperature approaches to ${T_{\mathrm{N}}}$, specifically 121 K for 2 L sample gained from the layer-dependent LD measurements (Fig. 2f), the corresponding LD signals dramatically drop until they vanish entirely above ${T_{\mathrm{N}}}$. Figure 4b shows the temperature-dependent LD intensity measured at the labelled 2 L position by the black dots in Fig. 4a, revealing a conspicuous LD sign change from positive to negative, signifying a magnetic domain flip. The orientation of the antiferromagnetic ordering before and after the domain flip is confirmed by conducting the angular-dependent LD measurements on the labelled 2 L position (Fig. 4c). We observe that the LD negative maximum changes from along 60$^{\circ}$ (-120$^{\circ}$) at 2 K to along 0$^{\circ}$ at 85 K, illustrating the transition from domain\_2 to domain\_1. Similarly, the left section of the 2 L flake change from domain\_2 to domain\_3, possibly originating from the domain wall motion of the domain\_3 from the adjacent 3 L flake. In contrast to the thermally-activated domain evolution in 2 L region, the magnetic domains in the 3 L region and the thick sample exhibit no change across various temperatures until they disappear above ${T_{\mathrm{N}}}$ (also see Fig. S2 for the thick sample). 

\hspace{1em}We attribute the origin of the magnetic domain evolution in the 2 L NiPS\textsubscript{3} to the competition between local strain and thermal fluctuation. Notably, we observe that upon conducting repeated LD mapping in various thermal cycles, the domain structure remains unchanged at 2 K, which implies that the domains are not degenerate but pinned by local effects such as strain\cite{FePS3domain2}. This notion is reinforced by the evidence that the residual LD signals at \textit{T} = 175 K (considerably higher than the ${T_{\mathrm{N}}}$ of bulk NiPS\textsubscript{3}) can depict the possible strain in the sample and match with the domain structure observed at 2 K (Fig. S11), which implies that the strain defines the domain distribution by perturbing the formation energy as illustrated in Fig. 4d. As for substantially thicker bulk samples, these local effects might be imperceptible in comparison to the stabilization of monoclinic stacking, resulting in the antiferromagnetic order along the crystalline \textit{a}-axis measured by magnetometry and neutron scattering\cite{NiPS3AFM}. In the case of atomically-thin NiPS\textsubscript{3} (e.g. 3 L sample), strain plays significant role to overcome the formation energy differences among the three domains, leading to the observation of the three-state nematicity (Fig. 4d). Nevertheless, the high stability of antiferromagnetism leads to the instant formation of magnetic domains restricted by strain at \textit{T} < ${T_{\mathrm{N}}}$, thus no domain flip is observed in 3 L and thicker NiPS\textsubscript{3}. However, when NiPS\textsubscript{3} goes down to 2 L, the spin fluctuation is stronger so the thermal fluctuation would overcome the formation energy differences of the domains, resulting in a non-stable domain distribution at high temperature (Fig. 4e). As the temperature drops considerably below ${T_{\mathrm{N}}}$, the thermal fluctuation becomes weak and allows the strain-defined domains to construct until reaching the stable domain distribution.

\section*{Conclusion}
In summary, through the measurements of magneto-optical effect combined with PL and Raman spectroscopy, we uncover the presence of the antiferromagnetism in bilayer and thicker NiPS\textsubscript{3}. Lower magnetic transition temperature (${T_{\mathrm{N}}}$) and larger critical exponent ($\beta_{\mathrm{LD}}$) are obtained in thinner NiPS\textsubscript{3} from the temperature-dependent LD measurements. The emergence of the three-state nematicity is observed in atomically-thin NiPS\textsubscript{3} down to 2 layer sample. We also report the thermally-activated domain evolution due to the strong spin fluctuation in 2 L NiPS\textsubscript{3}, imparting fundamental knowledge pivotal for future research endeavors including strategies for manipulating the antiferromagnetic behavior. Our study provides significant insights to the advancement of next-generation 2D antiferromagnetic spintronics.

\newpage

\section*{Methods}
\textbf{NiPS\textsubscript{3} Crystal Synthesis.} A stoichiometric ratio of elements (Ni:P:S $=$ 1:1:3, 1 g in total) and transport agent iodine (about 20 mg) were sealed into a quartz ampule with pressure pumped down to 1$\times$10\textsuperscript{$-$4} Torr. Then the ampule was heated in a two-zone furnace (650 $-$ 600 $^{\circ}$C) for 1 week and cooled down to room temperature. Bulk crystals were collected from the lower temperature end of the ampule. Thin flakes were obtained by mechanical exfoliation using scotch-tape and their thicknesses were determined by optical microscopy (Nikon DS-Ri2) and atomic force microscopy (Bruker Dimension 3000). Samples then were loaded into a closed-cycle optical cryostat by Montana Instruments or Quantum Design for all the temperature-dependent measurements.
\vspace{1\baselineskip}

\textbf{PL and Raman spectroscopy measurements.} PL and Raman measurements were conducted by using a con-focal microscope spectrometer (Horiba LabRAM HR Evolution) with a $\times$50 objective and 532 nm laser excitation. Signals were dispersed by a 600 gr/mm grating for PL, and by a 1800 gr/mm grating for Raman, detected with a liquid nitrogen cooled charge-coupled device (CCD) camera. The polarization angle-resolved PL measurements were performed by adding a linearly polarizer on the collective path. Co-linearly polarized Raman were conducted by adding a half-wave plate before the objective.
\vspace{1\baselineskip}

\textbf{Linear dichroism measurements.} A photo-elastic modulator (PEM; PEM-100, Hinds Instruments) was used on the incident path of the optical setup. A supercontinuum laser (NKT Photonics) resolved by a monochromater for 1.5 nm resolution was used as excitation, which was focused onto the sample using a $\times$50 objective. The back-scattered light was collected by the same objective and measured by an amplified photodiode (ThorLabs PDA100A2), in which the output was connected to a lock-in amplifier (Stanford Instruments SR865A) referenced to the second harmonic of the fundamental PEM frequency f $=$ 50 kHz. The beam incident on the PEM was prepared with linear polarization making an angle of 45$^{\circ}$ with respect to the PEM fast axis and the amplitude was modulated with a mechanical chopper at frequency of 578 Hz. The PEM retardance was set to 0.5\(\text{$\lambda$}\) to modulate the incident polarization. The total reflectance of the sample as a normalization, was monitored by a second lock-in amplifier referenced to the chopping frequency. A zero-order half-wave plate for changing the polarization states of the incident light after PEM was used and placed before the objective. All LD measurements except the LD spectra were conducted with the excitation wavelength of 633 nm.
\vspace{1\baselineskip}

\textbf{Polarized microscopy.} A broadband visible LED light, and linearly polarizers on both the input and output paths were used in a reflection geometry, with the images captured by a standard complementary metal-oxide-semiconductor-based monochrome camera (ThorLabs CS165MU).

\vspace{1\baselineskip}

\textbf{Data availability.} Data that support the study in this article are available from the corresponding authors on reasonable request.
\vspace{1\baselineskip}

\textbf{Acknowledgements.} This material is based upon work supported by the National Science Foundation (NSF) under Grant No. 1945364. H.G. acknowledges the support of BUnano fellowship from Boston University Nanotechnology Innovation Center. X.L. acknowledges the membership of the Photonics Center at Boston University. Work done by X.L. is also supported by the U.S. Department of Energy (DOE), Office of Science, Basic Energy Science (BES) under Award No. DE-SC0021064.
\vspace{1\baselineskip}

\textbf{Author contributions.} Q.T., C.A.O., R.C. and X.L. conceived the project. Q.T., H.G., and H.K. synthesized and characterized the NiPS\textsubscript{3} crystals. Q.T., C.A.O., and J.L. performed the optical measurements supervised by R.C. and X.L. Q.T. and C.A.O. performed the data analysis and interpretation. Q.T., C.A.O. and X.L. wrote the manuscript with assistance of all authors.
\vspace{1\baselineskip}

\textbf{Competing interests.} The authors declare no competing interests.

\newpage

\bibliography{references}

\newpage

\section*{Figures}
\begin{figure}[H]
\centering
\includegraphics[width=14cm]{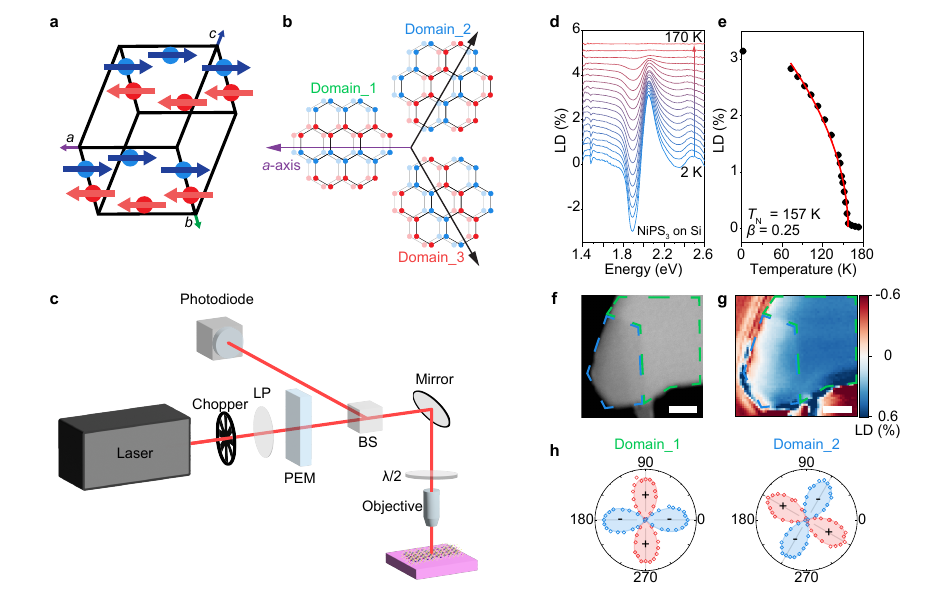}
\end{figure}
\textbf{Fig. 1. Magnetic structure and detection of NiPS\textsubscript{3}.} \textbf{a}, Schematic of the spin structure on Ni atoms of NiPS\textsubscript{3}. \textbf{b}, Three-state nematicity formed by zigzag antiferromagnetic orders. The spin direction up and down is represented by blue and red circles. \textbf{c}, Experimental setup for linear dichroism measurement. LP: linear polarizer, PEM: photo-elastic modulator, BS: beam splitter, $\lambda$/2: half-wave plate. \textbf{d}, Temperature-dependent LD spectra of a thick flake. \textbf{e}, Extracted LD values at the photo energy of 2.05 eV from \textbf{d}. The red line is the fitting using the formula $\mathrm{LD}(T) \sim\left|1-\frac{T}{T_{\mathrm{N}}}\right|^{2\beta_{\mathrm{LD}}}$, where ${T_{\mathrm{N}}}$ = 157 K, and $\beta_{\mathrm{LD}}$ = 0.25.  \textbf{f}, Polarized microscopy image and \textbf{g}, LD mapping of the same thick sample at 5 K, confirming the same two magnetic domains. Domain\_1 and domain\_2 are labelled by green and blue dashed lines, respectively. Scale bar: 5 {\textmu}m. \textbf{h}, Angular-dependent LD results of the thick sample on the two domain regions. 

\newpage
\begin{figure}[H]
\centering
\includegraphics {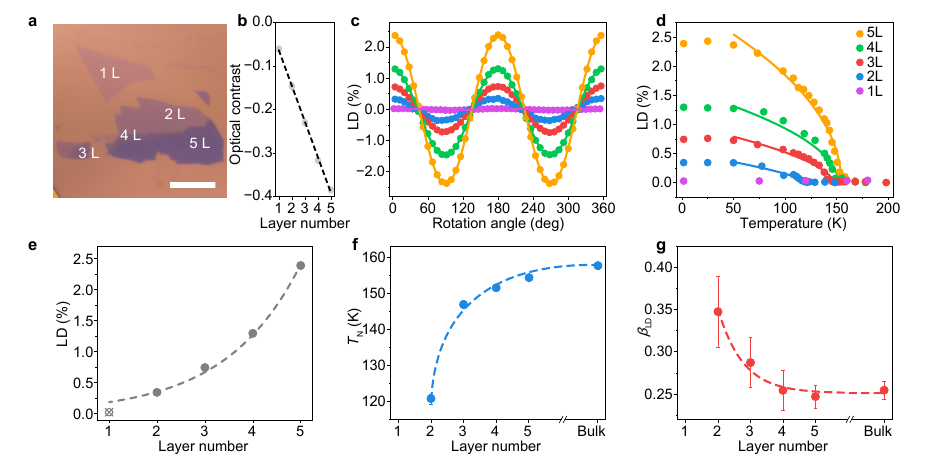}
\end{figure}
\textbf{Fig. 2. Thickness-dependent magnetism of NiPS\textsubscript{3}.} \textbf{a}, Optical image of exfoliated NiPS\textsubscript{3} flakes with the layer number indicated. Scale bar: 10 {\textmu}m. \textbf{b}, Optical contrast in red channel of the sample with different numbers of layers. \textbf{c}, Layer-dependent LD results measured at 1.8 K. \textbf{d}, Temperature-dependent LD results for 1-5 L NiPS\textsubscript{3} samples. Lines are the fitting using the formula, $\mathrm{LD}(T) \sim\left|1-\frac{T}{T_{\mathrm{N}}}\right|^{2\beta_{\mathrm{LD}}}$, where ${T_{\mathrm{N}}}$ is the transition temperature, and $\beta_{\mathrm{LD}}$ is the critical exponent. \textbf{e}, Layer-dependent LD intensity acquired from measurement at 1.8 K. \textbf{f}, Layer-dependent transition temperature, ${T_{\mathrm{N}}}$. \textbf{g}, Layer-dependent critical exponent, $\beta_{\mathrm{LD}}$. Dash lines in \textbf{e}, \textbf{f}, and \textbf{g} are guides to the eye.

\newpage
\begin{figure}[H]
\centering
\includegraphics {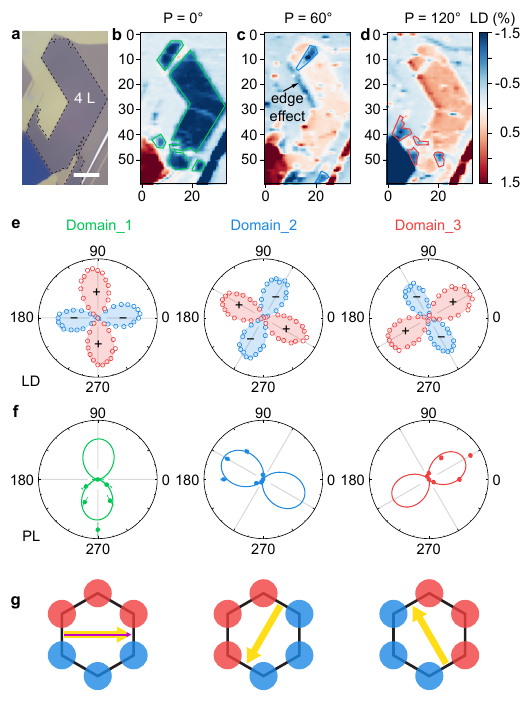}
\end{figure}
\textbf{Fig. 3. Three-state nematicity in a four-layer NiPS\textsubscript{3}.} \textbf{a}, Optical image of 4 L NiPS\textsubscript{3} flake outlined by black dashed lines. Scale bar: 10 {\textmu}m. \textbf{b-d}, LD intensity mapping on the same region in \textbf{a} with the incident laser polarization (P) at 0$^{\circ}$, 60$^{\circ}$ and 120$^{\circ}$. Domain\_1, domain\_2, domain\_3 are outlined by green, blue, and red lines, respectively. \textbf{e}, Angular-dependent LD results in 4 L NiPS\textsubscript{3} on three domain regions. \textbf{f}, PL intensity as a function of the collection polarization. \textbf{g}, Schematics of three magnetic domains formed by zigzag order. Red and blue circles represent opposite in-plane spin directions on Ni atoms. Yellow arrows and purple arrow indicate the N\'eel vectors and the \textit{a}-axis of the sample, respectively. All measurements are done at 4 K. 

\newpage
\begin{figure}[H]
\centering
\includegraphics[width=14cm] {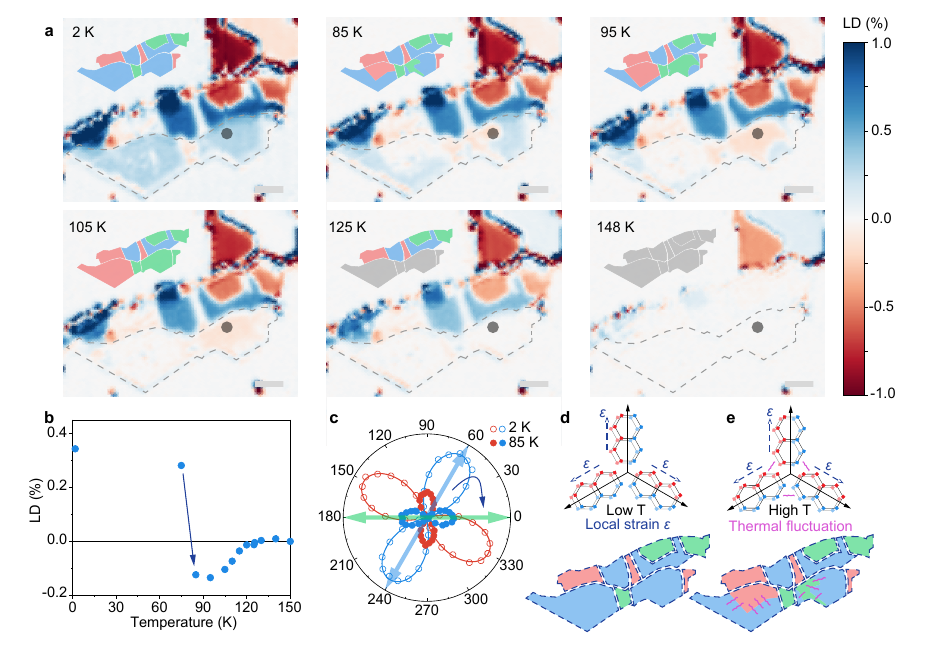}
\end{figure}
\textbf{Fig. 4. Magnetic domain evolution of NiPS\textsubscript{3} with varying temperature.} \textbf{a}, LD mapping images taken on an atomically-thin NiPS\textsubscript{3} flake from 2 K to 148 K. Signals measured at 175 K are subtracted as background. The inset cartoons picture the magnetic domain\_1, domain\_2, domain\_3 in green, blue, and red, respectively. When temperature is higher than ${T_{\mathrm{N}}}$, LD signals disappear and is represented in grey color. The dashed lines depict the 2 L NiPS\textsubscript{3} region. The labelled position by the black dots are tracked by angular-dependent LD measurements. Scale bar: 10 {\textmu}m. \textbf{b}, Temperature-dependent LD intensity for the labelled position of 2 L NiPS\textsubscript{3} in \textbf{a}. The sign change from positive to negative of LD intensity implies a magnetic domain flipping. \textbf{c}, Angular-dependent LD results on the labelled position in \textbf{a} at 2 K and 85 K. Green and blue arrows show the direction of the N\'eel vector. \textbf{d}, Mechanism of the domain evolution. Stable magnetic domain distribution confined by local strain is formed at low temperature. At high temperature, the thermal fluctuation provides sufficient energy for 2 L NiPS\textsubscript{3} to enable the domain flip.

\end{document}